\documentclass{article}

\usepackage{amssymb}
\usepackage{hyperref}
\usepackage{cleveref}
\usepackage{pdflscape}
\usepackage{soul}
\usepackage[square,numbers]{natbib}
\usepackage{graphicx}

\begin{document}

\title{IG Parser: A Software Package for the Encoding of Institutional Statements using the Institutional Grammar}

\author{Christopher Frantz\\Norwegian University of Science and Technology}

\date{}

\maketitle

\begin{abstract}
This article provides an overview of IG Parser, a software that facilitates qualitative content analysis of formal (e.g., legal) rules or informal (e.g., social) norms, and strategies (such as conventions) -- referred to as \emph{institutions} -- that govern social systems and operate configurally to describe \emph{institutional systems}. To this end, the IG Parser employs a distinctive syntax that ensures rigorous encoding of natural language, while automating the transformation into various formats that support the downstream analysis using diverse analytical techniques. The conceptual core of the IG Parser is an associated syntax, IG Script, that operationalizes the conceptual foundations of the Institutional Grammar, and more specifically Institutional Grammar 2.0, an analytical paradigm for institutional analysis. This article presents the IG Parser, including its conceptual foundations, the syntax specification of IG Script, and its architectural principles. This overview is augmented with selective illustrative examples that highlight its use and the associated benefits.
\end{abstract}

\noindent

\noindent

\section{Motivation and Background}

This article introduces IG Parser, a tool for the rigorous parsing of text in the form of institutional statements, which represent the fundamental unit of analysis of the Institutional Grammar (IG)~\cite{Crawford1995AInstitutions,FrantzSiddiki2022}, an analytical paradigm that approaches the analysis of institutional arrangements (i.e., structural description and behavioral regulation of governance regimes) expressed in spoken or written language based on a uniform syntax that is able to capture fine-granular institutional structure and semantic information embedded in so-called institutional statements. To support this process, this article performs a two-fold function, namely providing a \emph{detailed overview of IG Script}, the associated syntactic notation used for the operational coding, or as an intermediate structure, alongside the \emph{Parser}, i.e., the software that is able to parse this notation and generate analytically accessible output on that basis. 

Both the syntactic notation and the associated parser provide a starting point for a) facilitating systematic and rigorous encoding of institutional statements, and b) maintaining traceability from original input to encoded information using an output-agnostic intermediate representation. This hence addresses central key limitations of contemporary applications of the IG, namely a) promoting a \emph{methodologically rigorous application} in the encoding process (including traceability between original natural language and encoded statement), b) opening encoded data to a broader set of analytical techniques (e.g., statistical analysis, network analysis, modelling) by dissociating encoding from analysis based on an intermediate structure, as well as c) providing a basis to \emph{enable comparative studies using the IG}, an aspect that challenged due to the diverse representations and encoding strategies used for the IG.

The software is geared for the use both by non-technical users based on a user interface addressed at non-technical users that variably allows for entry based on key or touch input, as well as programmatic invocation based on an application programming interface (API) that consumes externally produced IG Script-encoded input and converts the statements into the desired output format. The modular nature of the software is open to the extension with additional output formats to serve specific analytical or downstream processing needs. The current UI exemplifies this by offering variants of tabular outputs, as well as a visual output structure. 

To date the software has found adoption in the context of IG Training workshops as a means to teach operational encoding while enabling low-threshold data exchange\footnote{Due to agnosticism about formatting of encoded information (e.g., no need for tabular structure, component order, or tab-based formatting), encoded content can easily be shared via chats and encoded in an ad hoc way in online settings.}, as well as initial research articles that draw on data encoded in the IG Parser. Finally, the syntax and examples in the companion book \cite{FrantzSiddiki2022} as well as Codebook \cite{FrantzSiddiki2020IGCodebook} are written in the IG Script notation and can be processed using the parser. Contemporary efforts of supporting the encoding are primarily focused on automation to facilitate analysis at scale \citep{Rice2020MachineGrammar,Heikkila2018a}, whereas the software introduced here follows the objective of promoting rigorous and consistent encoding, and thus supporting efforts toward standardization encoding in the first place, on which approaches to facilitate the automated encoding can build. 

In the following section, this article introduces the underlying syntactic notation IG Script that is at the core of the parser functionality. However, this introduction assumes basic conceptual understanding of the IG. Readers unacquainted with the IG, its features (e.g., components, regulative and constitutive statements, levels of expressiveness), applications and specific challenges motivating the development of the IG Parser, are recommended to consult Appendix \ref{app:background} prior to continuing.

\subsection{IG Script}
\label{sec:IGScript}

\emph{IG Script}, introduced in conjunction with the IG 2.0, provides an intuitively accessible notation that aligns with the structure of the original statement (i.e., it does not require ex ante reordering of institutional information to accommodate a specific input format), and, due to its inherent textual encoding, allows for efficient storage and transmission. IG Script differentiates between fundamental syntactic forms that allow the encoding of distinctive patterns relevant to capture institutional information comprehensively, and to accommodate the advanced granularity necessary to capture both low- and high-granular levels of expressiveness. The basic syntactic form relies on a component indication (e.g., $A$ for \emph{Attributes}, $I$ for \emph{Aim}, etc.), the full overview of which is provided in \Cref{tab:IgScriptSymbols}. 

\begin{table}[!h]
\centering
\begin{tabular}{|p{3.5cm}|l|}
\hline
\textbf{Symbol} & \textbf{IG 2.0 Component} \\
\hline
\multicolumn{2}{|c|}{\textit{Component symbols specific to regulative statements}} \\
\hline
A & Attributes \\
\hline
A,p & Attributes Property \\
\hline
D & Deontic \\
\hline
I & Aim \\
\hline
Bdir & Direct Object \\
\hline
Bdir,p & Direct Object Property \\
\hline
Bind & Indirect Object \\
\hline
Bind,p & Indirect Object Property \\
\hline
\multicolumn{2}{|c|}{\textit{Component symbols specific to constitutive statements}} \\
\hline
E & Constituted Entity \\
\hline
E,p & Constituted Entity Property \\
\hline
M & Modal \\
\hline
F & Constitutive Function \\
\hline
P & Constituting Property \\
\hline
P,p & Constituting Property Property \\
\hline
\multicolumn{2}{|c|}{\textit{Components shared across regulative and constitutive statements}} \\
\hline
Cac & Activation Condition \\
\hline
Cex & Execution Constraint \\
\hline
O & Or else \\
\hline
\end{tabular}
\caption{Component Symbols in IG Script}
\label{tab:IgScriptSymbols} 
\end{table}

Using the component symbols as a basis, we can engage in the encoding of distinctive statements at various levels of expressiveness highlighted in Tables \ref{tab:IgScriptSyntax1} and \ref{tab:IgScriptSyntax2}\footnote{In the upcoming Tables \ref{tab:IgScriptSyntax1} and \ref{tab:IgScriptSyntax2}, $cSymbol$ stands for component symbols -- the symbols listed in \Cref{tab:IgScriptSymbols}, and $logOperator$ reflects either of the operators `AND', `OR' and `XOR'.} and exemplified in the following. 

Whereas the simplest form of encoding merely indicates the component alongside the embedded value (e.g., ``\texttt{A(officer)}''), referred to as \emph{atomic component}, the following syntactic patterns show increasing levels of structural complexity. We will use the following running example to highlight those: 

\texttt{\small If officer observes or is made aware of violation, officer must fine and report violator to authority.}

Drawing the initial feature, the statement is encoded as 

\texttt{\small Cac(If officer observes or is made aware of violation), A(officer) D(must) I(fine and report) Bdir(violator) to Bind(authority).}

Atomic component encoding naturally only captures a very coarse-grained encoding. Extending the coding to the identification of logical component combinations (the resulting output of which would be multiple atomic institutional statements as discussed in the introduction), referred to as \emph{component combinations}, we arrive at the following: 

\texttt{\small Cac(If officer (observes [XOR] is made aware of) violation), A(officer) D(must) \textbf{I(fine [AND] report)} Bdir(violator) to Bind(authority).}

Recognizing nested structures embedded within individual components made up of institutional statement components themselves (component-level nesting), we can encode those as \emph{nested components}: 

\texttt{\small \textbf{Cac\{If A(officer) I(observes [XOR] is made aware of) Bdir(violation)\}}, A(officer) D(must) I(fine [AND] report) Bdir(violator) to Bind(authority).}

Assuming more complex embedded structures, such as multiple preconditions (\emph{nested component combinations}), exemplified using an expanded variant of the example statement, the syntax supports the following encoding: 

\texttt{\small \textbf{Cac\{Cac\{If A(officer) I(observes [XOR] is made aware of) \\ Bdir(violation)\} [AND] Cac\{if A(officer) I(deems) Bdir(intervention) Cex(safe)\}\}}, A(officer) D(must) I(fine [AND] report) Bdir(violator) to Bind(authority).}

Supporting the common use of compound component use (component pairs) in natural language, such as the combined use of action and objects, the syntax (exemplified in yet another variation of the working statement) supports this as \emph{component pair combinations}: 

\texttt{\small Cac\{Cac\{If A(officer) I(observes [XOR] is made aware of) \\Bdir(violation)\} [AND] Cac\{if A(officer) I(deems) Bdir(intervention) Cex(safe)\}\}, A(officer) D(must) \textbf{\{I(fine) Bdir(violator) [AND] I(file) Bdir(report) with Bind(district court)\}}.}

The final syntactic pattern focuses on extracting semantic features from statements -- whereas all the previous patterns focus on structure primarily. Whether for the purpose of disambiguating language, or to attach semantic labels, IG Script enables \emph{semantic annotations} by preceding the content with optional square brackets that hold either pre-defined, or user-defined labels.\footnote{Exemplary taxonomies for component- or statement-specific semantic annotations are provided in \cite{FrantzSiddiki2020IGCodebook} and \cite{FrantzSiddiki2022}.} Extending the previous statement accordingly, we can write

\texttt{\small Cac\{Cac[condition=violation]\{If A[role=enforcer](officer) I(observes [XOR] is made aware of) Bdir(violation)\} [AND] Cac[condition=safety]\{if \\A[role=enforcer](officer) I(deems) Bdir(intervention) Cex(safe)\}\}, \\A[role=enforcer](officer) D[stringency=high](must) \textbf{\{I[act=sanction](fine) \\Bdir(violator) [AND] I[act=report](file) Bdir(report) with \\Bind[act=authority](district court)\}} [statement-type=consequential].}

As showcased in this example, annotations attach the institutional function to individual components, such as annotating an activity such as `fine' as bearing the institutional function of `sanctioning' (see ``\texttt{\small I[act=sanction](fine)}''), but can also operate on nested expressions (see the activation conditions ``\texttt{\small Cac[condition=violation]}'' and ``\texttt{\small Cac[condition=safety]}''), combinations thereof, or entire statements (see annotation ``\texttt{\small [statement-type=consequential]}''). Annotations of statements or components can be partial or comprehensive, and are primarily oriented at analytical needs (e.g., reflecting conceptual linkage to theory or framework used for analysis, or logical expression for algorithmic treatment). Semantic annotations further operate cross-cutting and can be used in conjunction with any of the preceding patterns, ranging from atomic components to component pair combinations. 

Without further illustration at this stage, the same syntactic forms apply to constitutive statements, and further consider properties of individual components, such as ``\texttt{\small \dots~\textbf{\small Bdir,p(written)} Bdir(report) \dots}'', in which the distinctive features of a given actor, object, etc.~can be qualified in a fine-grained manner (e.g., to distinguish a written from a verbal report if analytically relevant). Any of the introduced patterns can further be combined with other patterns of the same or different kind (under indication of precedence where linked by different logical operators) to capture the structural complexity of statements comprehensively.\footnote{An overview of all IG Script features, beyond the ones shown here, is provided as part of the IG Parser documentation available under \url{https://github.com/chrfrantz/IG-Parser}.} Tables \ref{tab:IgScriptSyntax1} and \ref{tab:IgScriptSyntax2} summarize the central syntactic patterns supported by IG Script in order of increasing levels of expressiveness (from IG Core via IG Extended to IG Logico) and increasing syntactic complexity within those levels.\footnote{Legend: \texttt{\small cSymbol}: component symbol (see \Cref{tab:IgScriptSymbols} for symbols); \texttt{\small logOperator}: logical operator `AND', `OR', `XOR'); multiple patterns/examples are separated by semicolons.}

\begin{landscape}
\begin{table}[!h]
\footnotesize
\begin{tabular}{|p{4.5cm}|p{2cm}|p{6.5cm}|p{6.5cm}|}
\hline
\textbf{Syntactic Pattern} & \textbf{Level of Expressiveness} & \textbf{Pattern Description} & \textbf{Example/s (the underlined parts of the expression highlight the discussed feature)} \\
\hline
\texttt{cSymbol(content)} & IG Core & \textbf{Atomic component}: Component content at the basic level of granularity (i.e., cannot be semantically decomposed further) & \texttt{\ul{A(actor)}} \\
\hline
\texttt{cSymbol(content [logOperator] content)} & IG Core & \textbf{Component combination}: Combination of arbitrary number of atomic values linked via distinctive logical operators, with precedence indicates using parentheses. & \texttt{A(actor) D(may) \ul{I(fine [XOR] arrest)}}; \texttt{A(actor) D(must) \ul{I(monitor [AND] (fine [XOR] arrest))}} \\
\hline
\texttt{cSymbol\{ ... \}} & IG Extended & \textbf{Nested component}: Represents component-level nesting, where the component of concern is substituted by an institutional expression (e.g., institutional statement or state) consisting of atomic components. Nesting operates infinitely, and expressions can include any other encoding patterns. & \texttt{A(actor) D(must) I(act) under the condition that \ul{Cac\{A(actor) I(observes) Bdir(violation)\}}} \\
\hline
\texttt{cSymbol\{ cSymbol\{ ... \} [logOperator] cSymbol\{ ... \} ... \}} & IG Extended & \textbf{Nested component combination}: Reflects combinations of nested components of the same kind (e.g., combinations of activation conditions, combinations of execution constraints) within a given statement. These combinations can contain an arbitrary number of logical combinations, as long as precedence is indicated by corresponding braces. & \texttt{A(actor) D(must) I(act) under the condition that \ul{Cac\{Cac\{A(enforcer) I(observes) Bdir(violation)\} [AND] Cac\{ A(violator) I(attempts) Bdir(escape)\}\}}} \\
\hline
\texttt{\{ cSymbol\{ ... \} [logOperator] cSymbol\{ ... \} ... \}} & IG Extended & \textbf{Component pair combination}: Captures the one or more combinations of component pairs of any kind. In contrast to the previous case, component pairs are distinctively linked components such as action-object pairs in natural language (e.g., `\ul{pay rewards} or \ul{administer fines}'). Parsing of such component pair combinations result in an expansion of the institutional statement into multiple distinctive statements. & \texttt{A(enforcer) D(may) \ul{\{I(investigate) Bdir(compliance) [XOR]\\ I(delegate) Bdir(investigation) to Bind(colleague)\}}} \\
\hline
\end{tabular}
\caption{Syntactic Forms in IG Script (1/2)}
\label{tab:IgScriptSyntax1} 
\end{table}
\end{landscape}

\begin{landscape}
\begin{table}[!h]
\footnotesize
\begin{tabular}{|p{4.5cm}|p{2cm}|p{6.5cm}|p{6.5cm}|}
\hline
\textbf{Syntactic Pattern} & \textbf{Level of Expressiveness} & \textbf{Pattern Description} & \textbf{Example/s (the underlined parts of the expression highlight the discussed feature)} \\

\hline
\texttt{cSymbol[annotation]( ...~)} & IG Logico & \textbf{Semantic annotation (atomic components)}: Facilitates semantic or logical annotation of any encoding pattern (atomic component, nested components, any form of combinations, or statements) entirely based on pre-defined taxonomies/ontologies or study-specific annotation schemes. This syntax showcases the annotation of atomic components. & \texttt{A\ul{[role=enforcer]}(officer) D\ul{[stringency=high]}(must) I\ul{[act=sanction]}(fine) Bdir\ul{[role=target]}(violator)} \\
\hline
\texttt{cSymbol[annotation]\{ ...~\}} & IG Logico & \textbf{Semantic annotation (nested components)}: Facilitates semantic or logical annotation of any encoding pattern (atomic component, nested components, any form of combinations, or statements) entirely based on pre-defined taxonomies/ontologies or study-specific annotation schemes. This syntax showcases the annotation of nested components. & \texttt{...~Cac\ul{[condition=violation]}\{if A[role=violator](violator) I[act=violate](violates)\}} \\
\hline
\texttt{cSymbol[annotation]\{ cSymbol\{ ...~\} [logOperator] cSymbol\{ ...~\} ...~\}; cSymbol[annotation]\{ cSymbol[annotation]\{ ...~\} [logOperator] cSymbol[annotation]\{ ...~\} ...~\}} & IG Logico & \textbf{Semantic annotation (nested component combinations)}: Facilitates semantic or logical annotation of any encoding pattern (atomic component, nested components, any form of combinations, or statements) entirely based on pre-defined taxonomies/ontologies or study-specific annotation schemes. This syntax showcases the annotation of combined nested components. & \texttt{Cac\ul{[condition=observedViolation]}\{ Cac\ul{[condition=violation]}\{if A[role=violator](violator) I[act=violate](violates)\} [AND] Cac\ul{[condition=observation]}\{if A[role=monitor](monitor) I[act=observe](observes) Bdir(violation)\}\}} \\
\hline
\texttt{[annotation] cSymbol\{ ...~\} cSymbol\{ ...~\} ...; [annotation1] cSymbol\{ ...~\} cSymbol\{ ...~\} [annotation2] cSymbol\{ ...~\} ; } & IG Logico & \textbf{Semantic annotation (statements)}: Facilitates semantic or logical annotation of any encoding pattern (atomic component, nested components, any form of combinations, or statements) entirely based on pre-defined taxonomies/ontologies or study-specific annotation schemes. This syntax showcases the annotation of statements. Note that a statement can have an arbitrary number of statement-level annotations which occur in any position of the statement. & \texttt{\ul{[statement-type=consequence]} A[role=enforcer](officer) D[stringency=high](must) I[act=sanction](fine) \ul{[another statement-level annotation]} Bdir(violator), Cac[condition=violation]\{if A[role=violator](violator) I[act=violate](violates)\}} \\
\hline

\end{tabular}
\caption{Syntactic Forms in IG Script (2/2)}
\label{tab:IgScriptSyntax2} 
\end{table}
\end{landscape}

This syntax provides the basis for the IG Parser as a tool that consumes this format as generic input.  As part of the parsing process it generates a complex tree structure that can be transformed into diverse output formats, amenable to distinctive downstream processing needs -- the architecture and features of which are discussed in the following.

%\section{Software description}

\section{IG Parser: Architectural Overview}

\begin{figure}[h]
    \centering
    \includegraphics[width=1.0\textwidth]{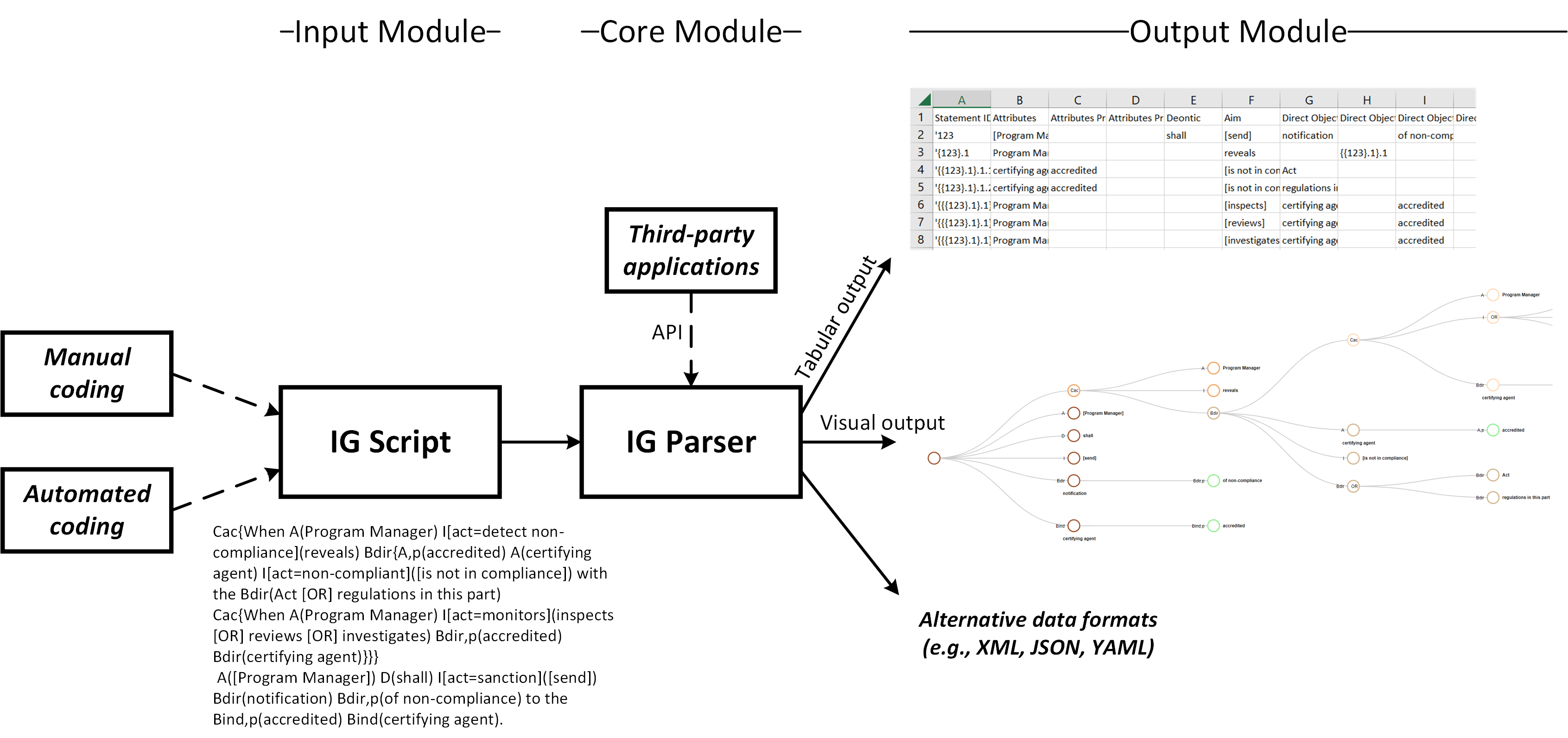}
    \caption{IG Parser Workflow}
    \label{fig:IGParserWorkflow}
\end{figure}

The parser itself consists of three main modules that separate input, parsing, and output generation, with the intent of offering conceptual openness for diverse input mechanisms (manual input, computational embedding via APIs), the core parsing module, and an extendable export module supporting selected exemplary output formats. The high-level operational workflow, showcasing the linkage between the input in the form of IG Script, the processing by the core parser module, as well as the varying output formats is schematically displayed in \Cref{fig:IGParserWorkflow}. 

\subsection{Input Module}

The current version of the \emph{Input Module} relies on a user interface (UI) that allows for manual input as well as specification and parameterization of desired output formats in which the results are provided. At this stage, two possible output formats are supported, including a tabular output that is primarily geared for downstream processing of decomposed statements using statistical programming (e.g., using R) or spreadsheet tools, as well as a visual output that displays the encoded statement as a tree structure to support the user in the encoding process, but also to derive additional analytical insights related to statement complexity. 

\subsubsection{Parameterizing Tabular Output}

Specific supported output formats include a comma-separated values (CSV) output\footnote{Per default, this output format uses the Pipe symbol (`$|$') as delimiter, since commas are tolerated as part of the input.}, as well as output in Google Sheets syntax to allow automated decomposition in the corresponding spreadsheet. This is augmented by a range of parameters, most notably the desired Statement ID that is used to generate Statement IDs for the output, and specifically sub-statement IDs, an aspect particularly relevant for generating decomposed \emph{atomic institutional statements} (see motivation in \Cref{sec:LevelsOfExpressiveness} and operational coding \Cref{sec:IGScript}). In addition, it supports the indication of the desired level of expressiveness that the output should reflect. Recall that the parser supports the downward compatibility of expressiveness, hence is able to provide output for statements encoded at higher levels of expressiveness (e.g., IG Logico) at lower levels of expressiveness (e.g., IG Extended or IG Core). %Finally, the tabular output parameterization supports convenience features such as the selective suppression of column headers and the inclusion of the IG Script-encoded statement in the generated output.

\subsubsection{Parameterizing Visual Output}

As an alternative to the tabular output, the visual output configuration likewise enables the parameterization of levels of expressiveness, as well as the selective in- or ex-clusion of showcasing property relationships associated with individual components. In addition, the ordering of selected components can be modified (activation conditions), to display the statement in the logical order of interpretation.\footnote{Activation conditions, for instance, can be displayed first, since these -- logically -- reflect the precondition of the applicability of the entire remaining statement.} Finally, the visual output allows for the generation of specific metrics (Degree of Variability) that measure the complexity of the institutional statement in terms of embedded action alternatives, acting as a proxy for the cognitive load of the involved statement (see \cite{FrantzSiddiki2022} for theoretical and conceptual foundations).

\subsection{Core Parser Module}

\begin{figure}[h]
    \centering
    \includegraphics[width=0.7\textwidth]{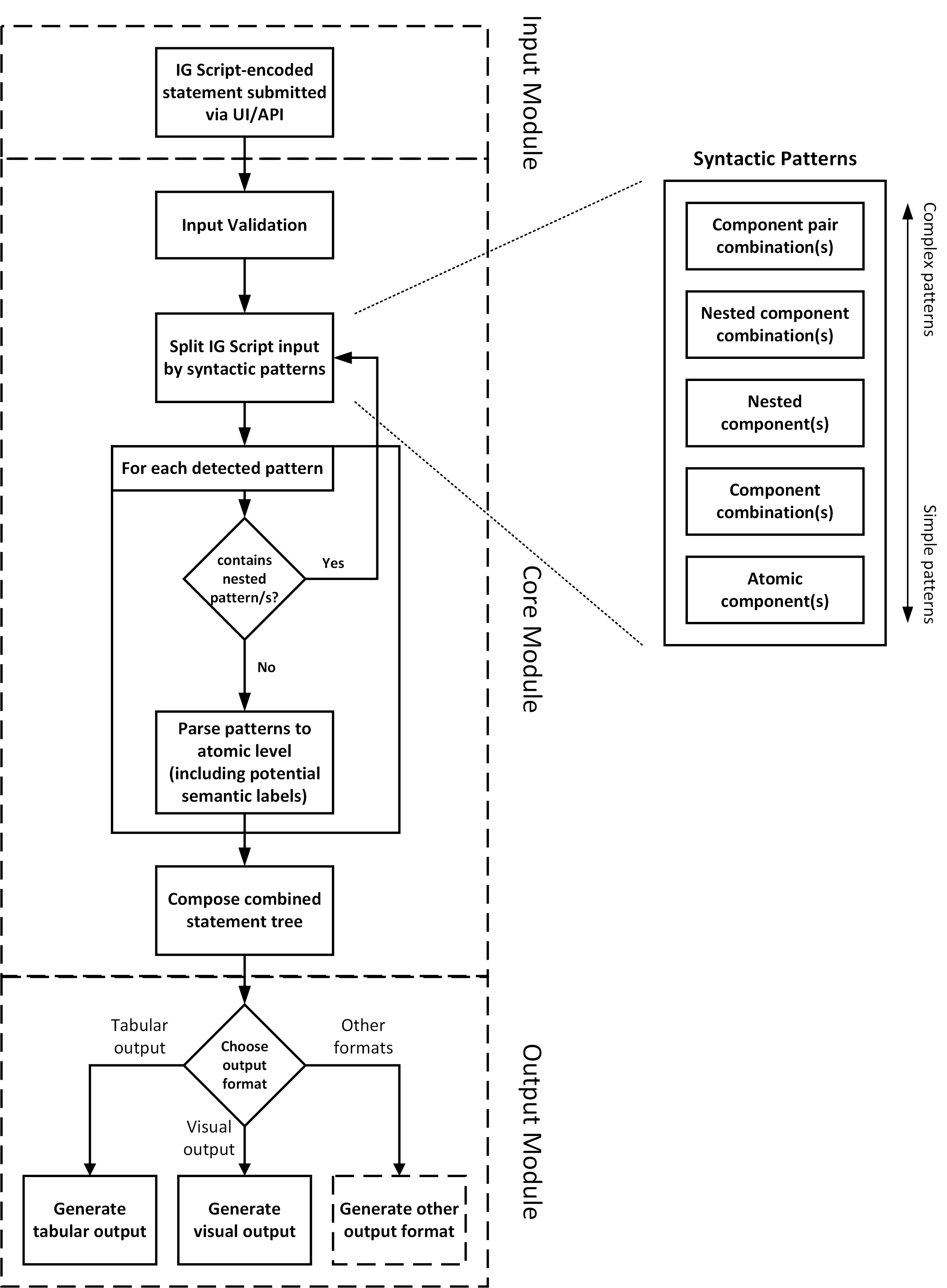}
    \caption{Parsing Process}
    \label{fig:ParsingProcess}
\end{figure}

The input provided as part of the Input Module in the IG Script syntax is the basis for the actual parsing of institutional statements as performed by the \emph{Core Parser Module}. It does so by following the process visualized in \Cref{fig:ParsingProcess} and explained in the following. 

The parser initially validates the input for syntactic correctness (e.g., complete brackets, indication of precedence of logical operators in the case of complex linkages), followed by scanning the entire input for embedded syntax patterns (highlighted in \Cref{sec:IGScript}) to assess the depth of encoding of the institutional statements, but also to provide the user with targeted feedback in the case of detected error in the input.

Following this, the individual patterns are individually processed in a recursive manner, in which the parser identifies nested patterns embedded within each of the identified top-level patterns, starting from the most complex to simplest pattern (see \Cref{fig:ParsingProcess}). It continues this process until all elements are parsed to an atomic level in which only individually encoded components remain, before incrementally constructing a statement tree based on pattern-specific processing to reflect the structural complexity of the input pattern comprehensively. Given the cross-cutting applicability of semantic annotations, those are processed independent of pattern-specific decomposition, since these are applicable across all patterns. The resulting tree structure represents the central internal representation of the institutional statement and provides the basis for generating distinct output formats produced by the \emph{Output Module}. While current use cases primarily focus on manual input based on the user interface, the Core Parser Module can alternatively be invoked via an Application Programming Interface (API) that enables programmatic access to parser functionality by third-party applications that offer their own facilities to provide generate parser input and process the generated response.

\subsection{Output Module}

The Output Module relies on the parsed institutional statement tree, as well as the parameterization provided by the Input Module in order to generate the output of interest. While supporting tabular and visual output explicitly at this stage, the architecture of the Output Module is conceptually open to support other output formats (e.g., distinctive file formats), supporting a modular extension.  

For the tabular output, the statements are decomposed into atomic statements, while tracking the logical linkages of the individual statements identified by generated Sub-Statement IDs (e.g., for component combinations). Where, for instance, the Statement ID provided as part of the input is `123', Sub-Statement IDs are `123.1', `123.2', etc. Where output for advanced levels of expressiveness such as IG Extended is requested (e.g., nested statements or higher-level patterns), the tabular output includes linkages for component-level nested statements based on ID references in the corresponding component field (e.g., an activation condition being decomposed into a separate atomic institutional statement provided in a separate Sub-statement). The nesting level can further be syntactically inferred based on the generated Sub-Statement ID (e.g., `\{123.1\}.1' is the first nested statement embedded in Sub-Statement `123.1'). As indicated before, the nesting can occur across multiple levels. An exemplary output reflecting this aspect is provided as part of the illustrative example showcased in \Cref{sec:Illustration}. 

As an alternative to the tabular output, the current version of IG Parser also generates a visual output. The visual output reflects the tree structure of the parsed institutional statement and, similar to the tabular output, relies on the parameterization via the \emph{Input Module} to configure the output. The user can interactively collapse and expand selected branches of the generated institutional statement tree\footnote{This feature is based on hierarchical tree structures produced using the visualization library D3.} to selectively explore the statement structure (e.g., to exclusively focus on activation conditions, action combinations, or actors), an aspect that both aids the understanding of the embedded structure, but also to use it as a basis to visually assess the coding of the statement as part of the encoding process. The user can interactively switch between both output formats, e.g., to use the visual output to support the encoding process -- a feature especially relevant for supporting the learning of novice coders, the encoding of particular complex institutional statements, or to support inter-coder reliability discussions. 

%\section{Illustrative examples}

\section{IG Parser: Illustrative Use}
\label{sec:Illustration}

In the following, we will illustrate the use based on selected institutional statements and for all of the showcased outputs, before discussing the impact of the IG Parser to date, opportunities for applications, as well as an outlook on upcoming extensions.

\subsection{Visual Output}

\begin{figure}[h]
    \centering
    \includegraphics[width=0.85\textwidth]{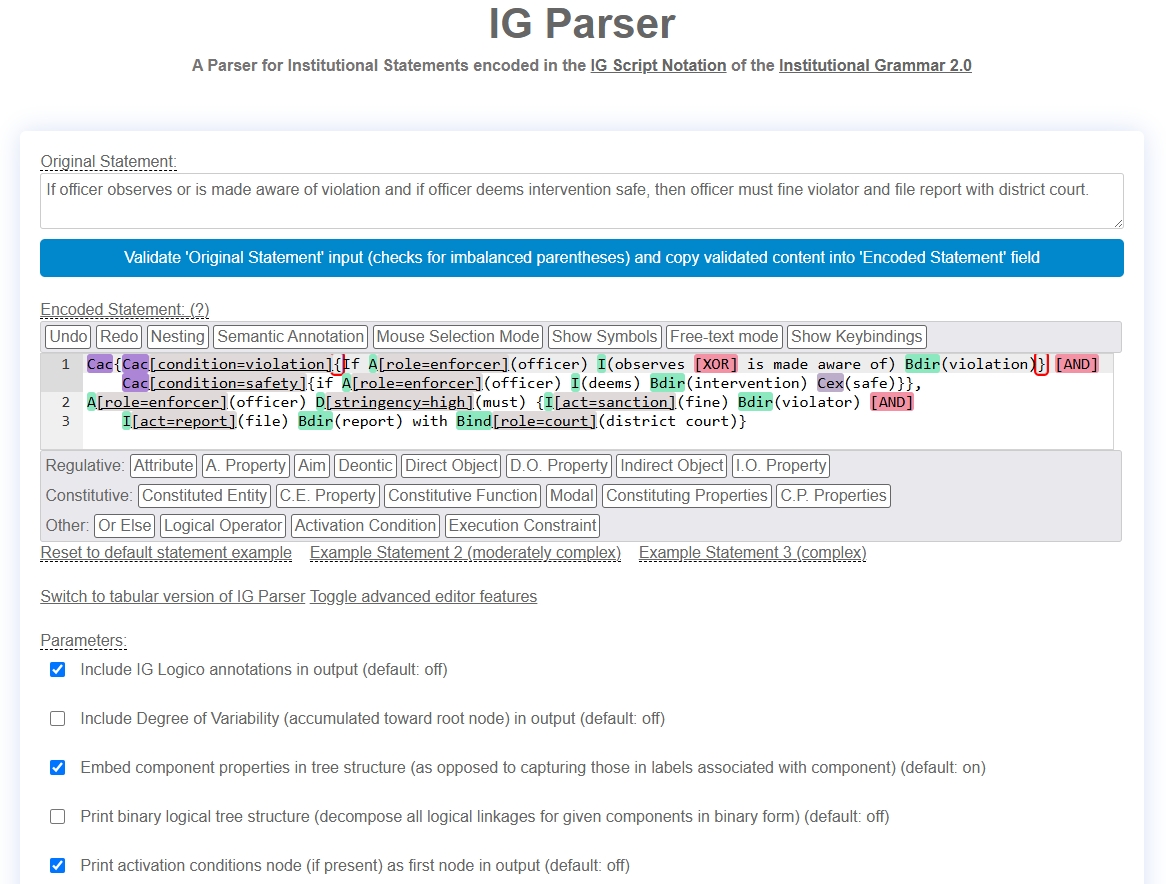}
    \caption{IG Parser User Interface for Visual Output}
    \label{fig:ParserUI}
    % Original Statement: 
    % If officer observes or is made aware of violation and if officer deems intervention safe, then officer must fine violator and file report with district court.
    % Encoded Statement: 
    % Cac{Cac[condition=violation]{If A[role=enforcer](officer) I(observes [XOR] is made aware of) Bdir(violation)} [AND] Cac[condition=safety]{if A[role=enforcer](officer) I(deems) Bdir(intervention) Cex(safe)}}, A[role=enforcer](officer) D[stringency=high](must) {I[act=sanction](fine) Bdir(violator) [AND] I[act=report](file) Bdir(report) with Bind[role=court](district court)}
\end{figure}

\Cref{fig:ParserUI} shows the user interface of the IG Parser (specifically, the visual output version), which allows the user to provide the original statement (for reference during the encoding process), as well as the entry field in which the encoding is performed. This entry field is augmented with additional features such as bracket matching to swiftly identify the scope of individual components, combinations and nested statements. Depending on the desired output, the parser provides corresponding input fields that parameterize the output generation. In the case of the visual output, this includes the ability to embed the semantic annotations in the output, selective suppression of component properties, enabling the restructuring of the output by displaying preconditions (activation conditions) prior to the remaining statement, as well as to configure output canvas size (to best accommodate the structural complexity of the resulting statement).

Following the generation of the output, the parser displays the result of the parsing process, or, whereas input parsing failed during validation, a corresponding error message that guides the user to specific issues related to the input. This is particularly relevant, since the parser operates on open input, which requires the handling of diverse input error scenarios. 

\begin{figure}[h]
    \centering
    \includegraphics[width=1.0\textwidth]{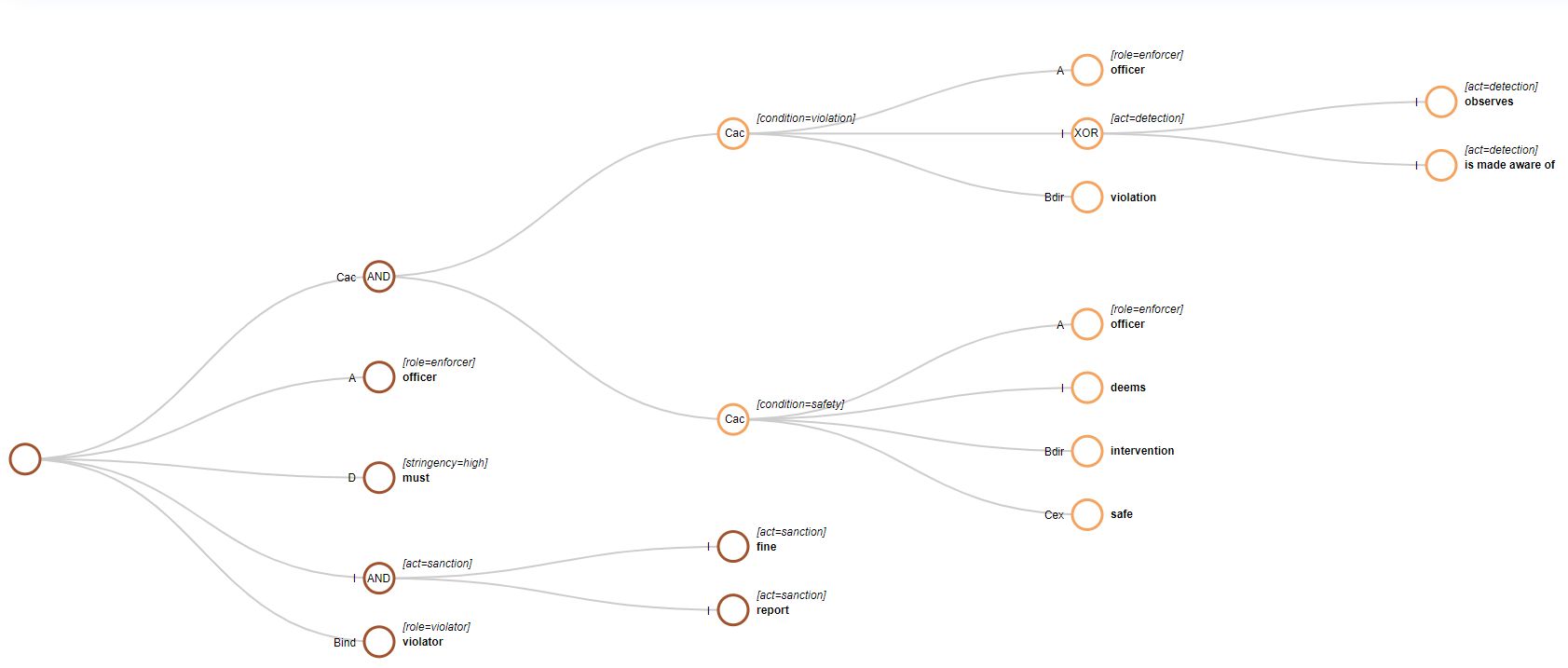}
    \caption{Generated Visual Output}
    \label{fig:ParserVisualOutput}
\end{figure}

The resulting visual output of the parsing process is shown below the input form as shown in \Cref{fig:ParserVisualOutput}. 

This output serves the inspection of the result, both to visually retrace the encoded statement, supporting an incremental encoding on the part of the user (IG Script is agnostic about non-encoded information), and to support the validation of the output as part of inter-coder reliability assessments. The visual output further supports an analytical function by providing information about the structural complexity of the encoded statement in terms of distinctive metrics specific to IG 2.0 (see Chapter 8 in \cite{FrantzSiddiki2022}). 

\subsection{Tabular Output}

\begin{figure}[h]
    \centering
    \includegraphics[width=0.95\textwidth]{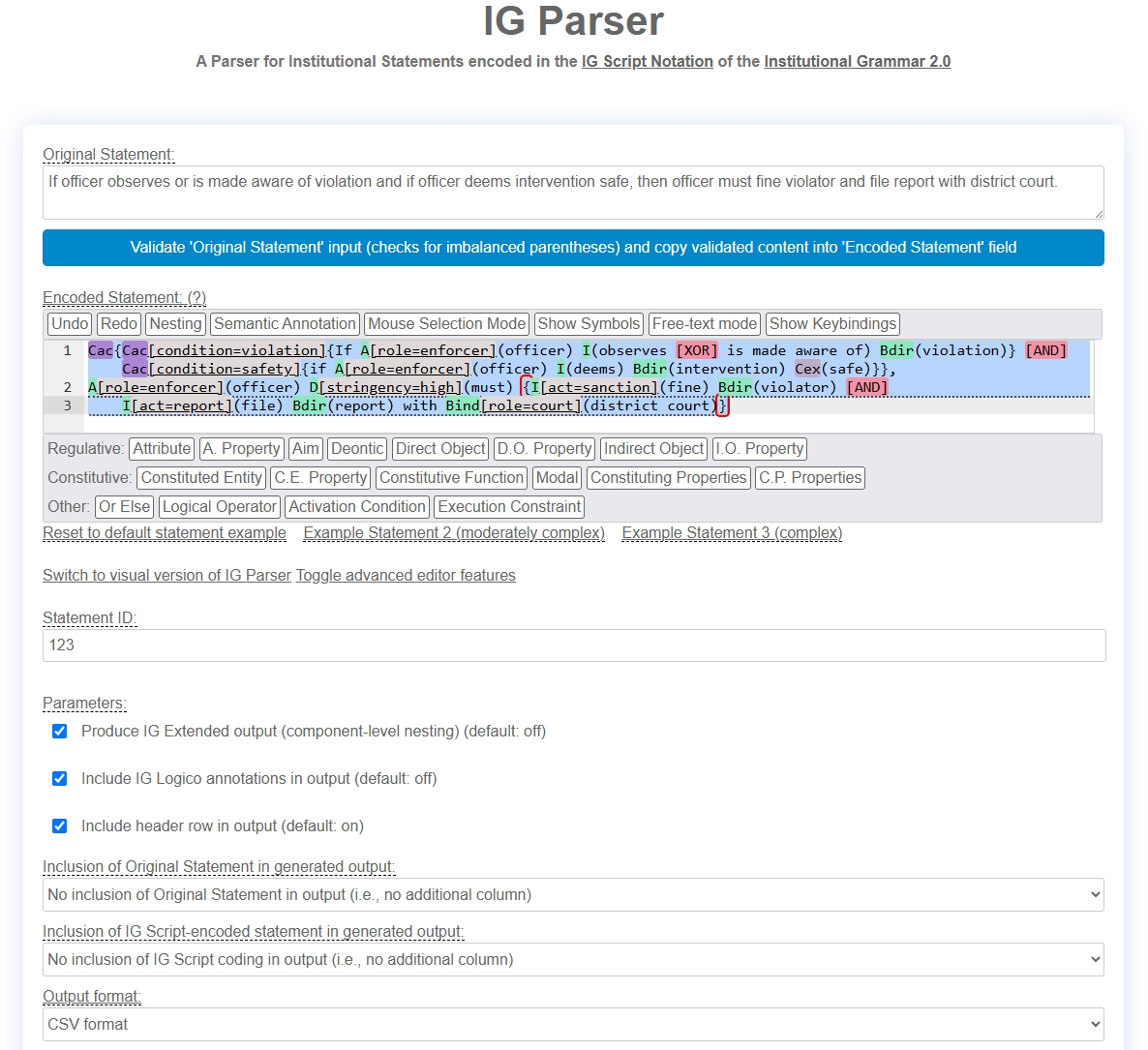}
    \caption{IG Parser User Interface for Tabular Output}
    \label{fig:ParserTabularUI}
\end{figure}

The tabular version of the IG Parser relies on the same input fields but varies with respect to output parameterization (as shown in \Cref{fig:ParserTabularUI}). Its primary purpose is to support downstream processing of larger numbers of statements in a tabular form. To this end, the IG Parser produces consistent output formats that allow the combination of multiple statements (e.g., an entire policy dataset) for analysis. Parameters include the specification of a Statement ID that the user wishes to use to identify the coded statement as part of her dataset, the choice as to whether the advanced nesting capability (component-level nesting and component combinations) are decomposed in the produced output (features associated with IG Extended as level of expressiveness), and optionally, the indication whether semantic annotations (a feature associated with IG Logico) are included in the generated output. To support the incremental coding of statements, column headers in the generated output can be selectively suppressed (e.g., only to show column headers for the first statement, since the structure of any following statement is identical). Where operating with the intent to process larger numbers of statement for downstream processing, users should ensure to use the same parameterization for all generated output. Per default, the UI saves the latest parameter set to support this by design.

\begin{figure}[h]
    \centering
    \includegraphics[width=1.0\textwidth]{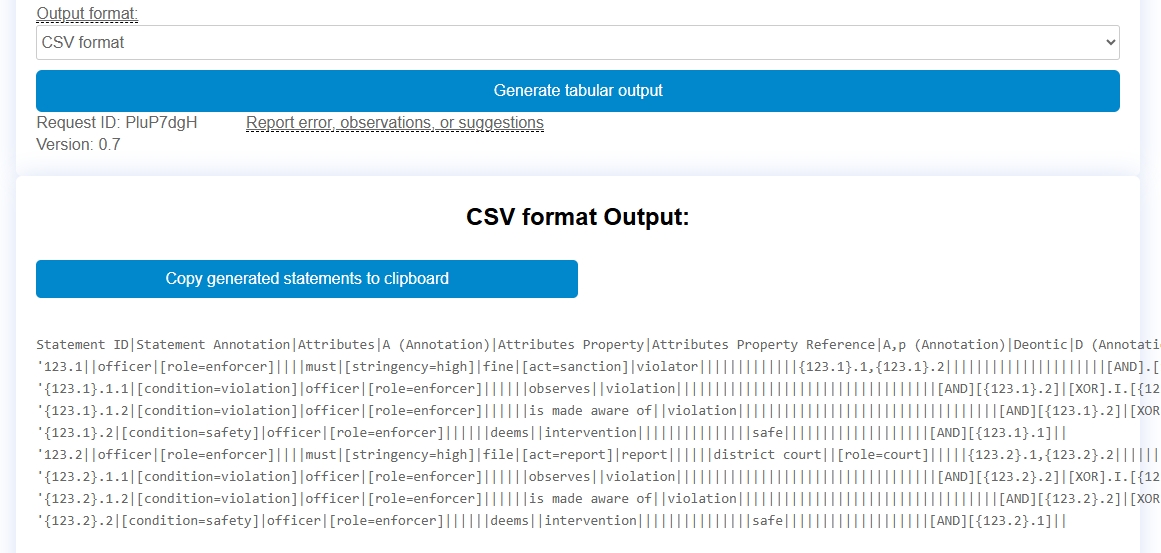}
    \caption{Generated Tabular Output}
    \label{fig:ParserTabularOutput}
\end{figure}

As indicated before, tabular output currently supports two output formats, namely CSV output drawing on the pipe symbol ($\mid$) as delimiter, or alternatively, Google Sheets output that can be directly processed in Google Sheets spreadsheets and automatically parses into the corresponding column layout without further mapping into columns (as is the case for the generic CSV format). To facilitate loss-less copying, the output features a button that copies the statement content to the clipboard for arbitrary external processing. \Cref{fig:ParserTabularOutput} exemplifies the generated output. 

To support novices during the encoding, the parser further features a set of example statements that can be selected in the UI, ranging from simple statements to very complex institutional statements. It further features an integrated help functionality that provides an overview of key syntactic patterns (introduced in \Cref{sec:IGScript} and \Cref{tab:IgScriptSymbols,tab:IgScriptSyntax1,tab:IgScriptSyntax2}).\footnote{All IG Script examples used in this article can be explored in the IG Parser under the URL \url{https://newinstitutionalgrammar.org/article-examples}.}

While not illustrated here, the features equally apply to regulative and constitutive statements, as well as combinations thereof (hybrid institutional statements), an aspect conceptually discussed elsewhere~\cite{FrantzSiddiki2022}.

\section{Impact and Outlook}

The IG Parser has been developed as a tool to make the feature set of the Institutional Grammar 2.0 operationally accessible for downstream processing. Based on the introduced IG Script syntax, the IG Parser provides the ability to process generic input in diverse output formats, allows the customization of the output based on desired feature set (e.g., levels of expressiveness), and is conceptually open for extension with prospectively useful output formats that enable novel analyses. As such, the parser enables policy analysts or institutional analysts more broadly, to encode and preserve institutional statements in an output-agnostic data format that offers greatest possible analytical flexibility. At the same time, the inspection features offered by the visual output allow for review and validation, while the output based on tabular format, for instance, ensures consistent and reliable output generation, a processing tasks that in most previous analyses drawing on the Institutional Grammar was handled manually, drawing to attention concerns about the reliability of encoding process. 

To date, the parser acts a companion to the conceptual foundations of IG 2.0 and has found use in various ongoing policy studies in the context of cybersecurity~\cite{Kianpour2025AnalysisGood}, socio-ecological systems and institutional complexity research.\footnote{Despite its focus on IG 2.0, the IG Parser can also be used for the encoding of institutional statements in the original Institutional Grammar syntax~\cite{Crawford1995AInstitutions} (sometimes referred to as IG 1.0) due to IG 2.0's backward compatibility.} In addition to generating output for the selective analyses presented in the companion book~\cite{FrantzSiddiki2022}, it is further used as part of -- at this stage four -- Institutional Grammar workshops throughout the past three years\footnote{See \url{https://newinstitutionalgrammar.org/events.html} for relevant training events.}, in which novices of broad disciplinary background are exposed to the opportunities of the Institutional Grammar more broadly, and the operational coding specifically, an aspect specifically supported by the visualization features of the parser. The various iterations of the tool have been presented at conferences relevant to the Institutional Grammar specifically, as well as institutional analysis more broadly. During this time, the parser has been incrementally refined based on feedback received from researchers and workshop participants, and, due to its inherent focus on processing unknown input, has been hardened with substantive testing.\footnote{The current version of the parser includes around 300 tests to test specific features as well as to capture typical cases of erroneous input.}

Specific research opportunities, beyond the statistical evaluation of the institutional information collected, include the analysis of institutional statements at greater structural depth based on the extended coding features introduced by the IG 2.0. This aspect, specifically, enables novel forms of complexity analysis on a statement level that previous work has not been able to address. The semantic annotations further allow for the encoding of logical and algorithmic information that enables the computational tractability of generated output (e.g., information about directionality of components or statements to support network analyses, symbolic representations of behavior to support formal or algorithmic treatment in specific programming languages or modelling frameworks). 

A practical opportunity lies in managing the trade-off between the potentially desirable retention of original language (e.g., to retain the proximate linkage between original and encoded statement) and preempting data cleansing needed to aggregate encoded content for quantitative analyses (e.g., for statistical or network analyses). A common challenge for such analyses is the disambiguation of language based on expressive diversity commonly found in regulation (e.g., `EU member' vs.~`Member' vs.~`Member states' vs. `Member state'). 

A forward-looking opportunity enabled by the IG Parser is to apply the enhanced feature set of the IG 2.0 to large-scale (large-n) studies based on the input format IG Script, which, due to its consistent structural patterns, is particularly amenable for automated generation of encoded institutional statements or their semantic labelling based on NLP techniques (e.g., Named Entry Recognition, Semantic Role Labelling) as well as pre-trained large-language language models. Specifically, the effort associated with encoding fine-grained institutional statements has to date been a factor that has restricted most IG studies to relatively small sample sets. The reliability associated with systematic guidance for the encoding and deterministic processing further drives methodological rigor, an aspect central for performing robust comparative studies in the long run (e.g., cross-domain comparison of policies or behavior), and hence enhance analytical opportunities in the research field of computational institutional science. 

Ongoing and future extensions of the parser focus on improving its practical utility, including ongoing efforts to enable corpus management (i.e., the management of complete datasets including batch processing and support of collaboration functionality), as well as exploring alternative UIs to facilitate diverse forms of input encoding (e.g., visual input) and output generation (e.g., domain-specific formats or data exchange formats such as UIMA CAS\footnote{See \url{https://www.oasis-open.org/committees/documents.php?wg_abbrev=uima}}) as amenable to different, heterogeneous communities and disciplines engaged in institutional analysis. This is enabled by the ability to variably invoke the parser as an API, or its embedding as a software module in domain-specific applications or entire analytical tool chains. 

IG Script as a notation and IG Parser as the corresponding parser are candidates to further harmonize the syntactic and semantic encoding of institutional statements to ensure methodological rigor in the area of IG-based institutional analysis, while enabling novel research opportunities by making the choice of structural features and divergent (or compatible) conceptual interpretations explicit and hence serve as a uniform structural interface between natural language and computationally accessible representation, while striving toward aligned practices in the encoding and data representation across research efforts engaging institutional analysis.

\bibliographystyle{elsarticle-num} 
\bibliography{references.bib}

\newpage

\appendix{}
\section*{Appendix}

\section{Current code version}
\label{lbl:codeVersion}

\begin{table}[!h]
\begin{tabular}{|l|p{5cm}|p{8cm}|}
\hline
\textbf{Nr.} & \textbf{Code metadata description} & \textbf{Metadata} \\
\hline
C1 & Current code version & v0.7 \\
\hline
C2 & Permanent link to code/repository & https://github.com/chrfrantz/IG-Parser \\
\hline
C3  & Deployed version & Link to deployed executable: 

https://ig-parser.newinstitutionalgrammar.org/ \\
\hline
C4 & Legal Code License   & GPL \\
\hline
C5 & Code versioning system used & git \\
\hline
C6 & Software code languages, tools, and services used & Go, Javascript, docker, docker compose \\
\hline
C7 & Compilation requirements, operating environments \& dependencies & The software runs on any operating system supported by Go.

Compilation requirements: 
\begin{itemize}
    \item Go 1.16 or higher (golang.org) 
\end{itemize}
Dependencies (shipped with code base): 
\begin{itemize}
    \item Ace (github.com/ajaxorg/ace)
    \item D3 (d3js.org)
\end{itemize} \\
\hline
C8 & Link to developer documentation/manual & https://github.com/chrfrantz/IG-Parser/blob/main/README.md \\
\hline
C9 & Support email for questions & christopher.frantz@ntnu.no \\
\hline
\end{tabular}
%\caption{Code metadata (mandatory)}
%\label{} 
\end{table}

%\appendix{Appendix}

\section{Conceptual Foundations of the Institutional Grammar}
\label{app:background}

\subsection{Institutions}

Understanding the functioning of social systems, whether at group, organizational, societal, or international level, relies on the ability to extract and analyze the rules that coordinate behavior, mitigate (or sometimes provoke) conflict and drive the necessary cohesion in diverse and open societies. Those \emph{institutions} can take various forms, manifesting as observed behavioral patterns of collective action, and of course in written form, be it in the form of instructions, regulations, laws, or other form of private or public policy. One way of making such -- typically qualitative -- information analytically accessible is to employ content analysis techniques. While text analysis is well established for capturing features of natural language (e.g., based on dependency tree parsing in the context of Natural Language Processing), language used for institutions is distinct, since a) not all language expresses rules, and b) institutional language displays distinctive functional properties that can be expressed in a uniform syntax whose features correspond to distinctive semantic features of rules. One approach to capture the essential features of such language at a fine-grained level is the \emph{Institutional Grammar}~\cite{Crawford1995AInstitutions,FrantzSiddiki2022}, a paradigm for institutional analysis that has found broad and long-standing disciplinary application, covering areas such as legal analysis~\cite{DeMattee2023AInstitutions}, political science, specifically public policy and administration~\cite{Siddiki2014AssessingStates,Stupak2020TheRules,Pielinski2022Keeping19892014,Chen2023ComparingPolicies,Kianpour2024AnalysisGood},  socio-ecological systems analysis~\cite{Lien2020TheResearch}, as well as computational social scientists interested in the study of institutions~\cite{Smajgl2010,Frantz2015c,Ghorbani2016ManagingAction}.\footnote{For comprehensive reviews of the application landscape, please consult \cite{Pieper2023TheReview} and \cite{Siddiki2022InstitutionalGrammar}.}

\subsection{Institutional Grammar}

The \emph{Institutional Grammar (IG)}, originally devised by Crawford and Ostrom~\cite{Crawford1995AInstitutions,Crawford2005}, and subsequently revised and extended by Frantz and Siddiki~\cite{FrantzSiddiki2022}, represents such `institutional language' in terms of institutional statements, where \emph{institutional statements} ``describe actions for
actors within particular contexts, or parameterize features of an institutional
system within particular contexts.''~\cite{FrantzSiddiki2022} Such statements can variably take the form of \emph{regulative institutional statements} that describe expected or permitted behavior for actors, and \emph{constitutive statements} that describe features of an institutional setting (institutional facts or acts), including the definition of actors, actions, venues, as well as status specifications (e.g., rights). Both forms of institutional statements are comprised of components that enable the fine-granular encoding of distinctive functional aspects. 

Reiterating the institution characterization offered in the introduction, while commonly applied to legal texts, institutional statements are intended to capture any regulating content irrespective of context (e.g., public, private), origin (e.g., formal decision, informal agreement) or form (e.g., written, spoken, observed). Combined, such statements afford the analysis of structure and behavior within \emph{institutional systems}. 

\subsection{Regulative Statements}

Regulative statements are composed of the following components (according with their function): 

\begin{itemize}
\item \emph{Attributes} -- Individual or corporate actor/s who is/are permitted, expected (or not expected) to carry out the action regulated in the statement.

\item \emph{Deontic} -- A prescriptive or permissive operator that defines the extent to which the action described in the institutional statement is compelled, restrained, or permitted.

\item \emph{Aim} -- The action or outcome regulated in the statement.

\item \emph{Object} -- The receiver/s or target/s which the action is directed to. The Object is differentiated into \emph{direct} (i.e., animate or inanimate object to which the action is directly applied) and \emph{indirect object} (i.e., object that is affected by the action application to the direct object).

\item \emph{Context} -- Captures the conditions under which the statement applies (\emph{activation condition}), as well as qualifications or moderation of actions in execution (\emph{execution constraint})

\item \emph{Or else} -- The consequence associated with non-compliance or -fulfillment of the regulated action. This consequence is represented as a separate linked institutional statement compromised of the same components mentioned above.

\end{itemize}

Of those components only \emph{Attributes}, \emph{Aim} and \emph{Context} components are required for any institutional statement\footnote{Note that the \emph{Context} component can be implied. In the absence of an explicit specification it defaults to the activation condition `under any condition' and the execution constraint `without constraints'.}; the \emph{Object}, \emph{Deontic} and \emph{Or else} components are optional. 

To illustrate the operational coding of regulative statements, we will rely on the following statement: `Drivers must stop their vehicle in front of the stop line when the traffic light is red.' 

This statement is encoded as follows: 

\begin{itemize}
    \item \emph{Attributes}: Drivers
    \item \emph{Deontic}: must
    \item \emph{Aim}: halt
    \item \emph{Direct Object}: their vehicle
    \item \emph{Execution Constraint}: in front of the stop line
    \item \emph{Activation Condition}: when the traffic light is red
\end{itemize}

\subsection{Constitutive Statements}

Constitutive statements -- statements parameterizing an institutional setting by constituting, modifying or otherwise affecting entities of institutional relevance -- are encoded using the following structure:

\begin{itemize}
\item \emph{Constituted Entity} -- Entity being constituted, reconstituted or otherwise modified by an institutional statement where the nature of the constitution is captured in the constitutive function 

\item \emph{Modal} -- An operator signaling necessity or possibility of the constitution captured in the institutional statement

\item \emph{Constitutive Function} -- The verb describing the nature of the constitution expressed in the institutional statement and resulting in the constituted entity. A potential constituting property serves as input to the constitutive function.  

\item \emph{Constituting Properties} -- Parameterizes the constituted entity via the constitutive function

\item \emph{Context} -- Captures the conditions under which the statement applies (\emph{activation condition}), as well as qualifications or moderation of the constitution process (\emph{execution constraint})

\item \emph{Or else} -- The consequence associated with the non-satisfaction of the statement. This consequence is represented as a separate linked institutional statement (consequential statement) of regulative or constitutive kind and compromised of the corresponding components. In contrast to consequential regulative statements that reflect social consequences (e.g., sanctions such as punishments), the consequence of non-fulfillment of a constitutive statement is commonly of existential kind (e.g., indicating the invalidity of the underlying policy).

\end{itemize}

For constitutive statements, the \emph{Constituted Entity}, \emph{Constitutive Function} and \emph{Context} components are compulsory\footnote{As with regulative statements, the \emph{Context} component carries implicit default values.}, with all other being optional -- similar to regulative statements. 

Exemplifying constitutive statements using the example `Traffic lights are sets of red, amber, and green lights at the places where roads meet'\footnote{This example is intentionally borrowed from Collins Dictionary to showcase the general applicability of the IG to any constitutive statement, irrespective of form and origin.}, the coding is as follows: 

\begin{itemize}
    \item \emph{Constituted Entity}: Traffic lights
    \item \emph{Modal}:\footnote{Absent explicit specification, the Modal value is implied as `necessary', i.e., traffic lights must \emph{necessarily} mean sets of red, amber and green lights \dots}
    \item \emph{Constitutive Function}: are
    \item \emph{Constituting Properties}: sets of red, amber, and green lights at the places where roads meet
    \item \emph{Context}:\footnote{As in the previous case, absent explicit specification of activation conditions and execution constraints their respective values default to `under any circumstance' and `without constraints'.}
    \item \emph{Or else}:\footnote{In this example, the consequence of non-fulfillment of the statement implies the non-existence of traffic lights in the context of the institutional setting (irrespective of their real-world presence).}
\end{itemize}

\subsection{Levels of Expressiveness}
\label{sec:LevelsOfExpressiveness}

The IG is devised to be amenable to wide range of analytical techniques that are able to process different levels of granularity in input information, variably focusing on structural depth as well as added semantic information. The IG 2.0 organizes these features by levels of expressiveness~\cite{FrantzSiddiki2022} that distinguish between a basic structural coding (\emph{IG Core}), an extended structural coding that recognizes complex nested structures (\emph{IG Extended}) and a semantic level that allows additional annotation of semantic information as well as affording basic logical transformations (\emph{IG Logico}). 

Examples of these structural extensions include the combination of components of the same type (e.g., `divide AND conquer'), but also include the nesting of complete statements within components of others. A typical example are preconditions that follow the same component structure as the main institutional statement (e.g., `If drivers drive on public roads, drivers must stop \dots', where the leading part reflects actor, action and context, followed by the original example statement). Implied in this nesting is the ability to combine constitutive and regulative institutional statements (e.g., the definition of a specific concept may be the precondition for a regulative statement to apply in the first place). Semantic extensions, a feature associated with the level of expressiveness IG Logico, refer to the ability to superimpose semantics relevant for the analysis that may be devised by the researcher, or be theoretically or empirically derived. One example of deriving semantics from theory is the classification of behavior as compliance (or variably violating behavior). Drawing on the earlier example, stopping the vehicle at the red light is -- from an institutional standpoint -- interpreted as a compliance signal; driving over a red light would be the corresponding violation. 

\subsection{Challenges}

A fundamental challenge with the processing of statements of such complexity (let alone variations in complexity based on the different levels of expressiveness) is the reliable encoding into a tree structure that, depending on chosen output, decomposes compound institutional statements into \emph{atomic institutional statements}. Taking, for instance, a typical expression such as ``Officers must issue warning or fine violating drivers.'', this functionally reflects two obligations, namely the discretion to either \emph{issue a warning}, or to \emph{fine} a violating driver. The resulting decomposed institutional instructions are logically-linked \emph{atomic institutional statements}. For analytical treatment a single institutional statement may thus be decomposed into multiple atomic statements, reflecting the most common challenge in downstream processing, but also highlighting the distinction between a sentence in policy text or spoken language, and an atomic institutional expression, for instance to assess the complexity of such expression based on the number of resulting atomic institutional statements, or to assess compliance (or non-compliance) with individual instructions as part of a computational analysis. More complex forms of such combinations, such as combinations of preconditions (activation conditions) or combinations of component pairs are discussed in the context of the encoding syntax IG Script (see \Cref{sec:IGScript}). that facilitates the extraction of such structural complexity.

The central purpose of the IG Parser is to address the overarching challenge of affording a semi-automated and reliable encoding of institutional statements of arbitrary complexity for diverse analytical purposes.

To this end, the IG Parser is able to capture all these distinctive features of the IG, while, at the same time, a) \emph{ensuring rigorous encoding of institutional content based on a uniform syntax}, b) maintaining \emph{proximate readability and portability of encoded information}, c) \emph{maintaining backward compatibility of levels of expressiveness} (i.e., information encoded at higher levels of expressiveness can be output for analyses that only operate on features of lower levels of expressiveness, hence enabling tailored output for distinct analytical opportunities, while relying on the same input), and d) displaying flexibility by \emph{accommodating of diverse existing output formats and being conceptually open to unknown out- and input formats} (e.g., novel user interfaces, application programming interfaces). 

To integrate all these features, as part of the encoding process the parser relies on a syntactic notation IG Script that makes the conceptual features of IG 2.0 accessible for analytical processing, which is introduced as part of the main text.

\end{document}